\documentclass[reprint,amssymb, amsmath,pre]{revtex4-1}
\usepackage{docs}%
\usepackage{graphicx}%
\usepackage{bm}%
\usepackage[colorlinks=true,linkcolor=blue]{hyperref}%
\expandafter\ifx\csname package@font\endcsname\relax\else
 \expandafter\expandafter
 \expandafter\usepackage
 \expandafter\expandafter
 \expandafter{\csname package@font\endcsname}%
\fi
\begin{document}

\title{Effect of a sub and suprathreshold periodic forcing an excitable glow discharge plasma near its bifurcation point.}%

\author{Md. Nurujjaman}%
\email{jaman\_nonlinear@yahoo.co.in}
\affiliation{Tata Institute of Fundamental Research Centre, TIFR Centre for Applicable Mathematics,
Post Bag No: 6503, Sharada Nagar, Chikkabommasandra, Bangalore 560065, India, and Plasma Physics
Division, Saha Institute of Nuclear Physics, 1/AF, Bidhannagar, Kolkata -700064, India.}%

\author{A. N. Sekar Iyengar}%
\email{ansekar.iyengar@saha.ac.in}
\affiliation{Plasma Physics Division,
Saha Institute of Nuclear Physics,
1/AF, Bidhannagar,
Kolkata -700064, India.}%

\date{March 2010}%
\revised{\today}%

\begin{abstract}
In this paper nonlinear dynamics of a periodically forced excitable glow discharge plasma has been studied. The experiments were performed in glow discharge plasma where excitability was achieved for suitable discharge voltage and gas pressure. The plasma was first perturbed
by a subthreshold sawtooth periodic signal, and we obtained small subthreshold oscillations. These oscillations showed resonance when the frequency of the perturbation was around the characteristic frequency the plasma, and hence may be useful to estimate characteristic of an excitable system. On the other hand, when the perturbation was suprathreshold, system showed frequency entrainments.  We obtained harmonic frequency entrainments for perturbation frequency greater than the characteristic frequency of the system and for lesser than the characteristic frequency, system showed only excitable behavior.
\end{abstract}

\maketitle

\section{Introduction}

Dynamics of a nonlinear plasmas under an external periodic perturbations have been studied extensively in several plasma systems, where different nonlinear phenomena, like periodic pulling, frequency entrainment, etc., were observed~\cite{prl:abrahams,physcript:Klinger,pop:dinklage,pop:Rohde,ppcf:Gyergyek,pop:Koepke,pla:Klinger,pop:Klinger,pre:klinger}. For suitable plasma parameters these systems may also show excitable dynamics~\cite{pre:jaman,pop:dinklage,prl:LinI}. Excitable plasmas show noise-induced resonances under both stochastic and periodic perturbations~\cite{pre:jaman,pop:dinklage,prl:LinI} that have  also been observed in many other physical, chemical, biological and electronics systems ~\cite{prer:jaman,prl:gang,prl:wiesenfeld,pre:strogatz,prl:pikovsky,prl:locher,pre:bascones,rmp:sagues,pre:dubbeldam,prl:stock,physrep:linder,pre:santos1,pre:lindner,prl:Giacomelli,pre:lee,pre:miyakawa}. Though excitable systems show rich dynamical behavior under a pure periodic forcing, little study is carried out in this area~\cite{pre:Mendez,pre:larotonda}. In this paper, our aim to explore the dynamics of glow discharge plasma near excitable region due to periodic perturbations.

The basic characteristic of an excitable system is that it shows a fixed point or coherent limit cycle oscillations depending on the value of the control parameter (CP) of the dynamics. The point where the system changes from oscillatory to fixed point behavior is called threshold or bifurcation point.   Now if the system shows fixed point behavior for the value of CP below the threshold, then the dynamics will be limit cycle oscillations on the other side of the threshold or vice versa, depending upon the system properties. Response of an excitable system at fixed point state to an external perturbation applied on the CP depends on the perturbation amplitude. When the perturbation is subthreshold, i.e., the amplitude is small such that the threshold is not crossed, the system shows small oscillations around the fixed point~\cite{physrep:linder}. These subthreshold or small oscillations may bear the characteristic of system dynamics~\cite{prl:kaplan,J.theor.Biol.:clay,r.soc:masoller}. On the other hand, when the perturbation is suprathreshold, i.e., it's amplitude is large enough to cross the threshold, the system returns to its fixed point deterministically, i.e., once the threshold is crossed, the system becomes almost independent of the perturbation and comes to its fixed point state traversing one limit cycle~\cite{prl:gang,prl:wiesenfeld,prl:Giacomelli,pre:strogatz,chaos:Coullet}. In plasma, excitability may be obtained through Hopf~\cite{prl:LinI,pop:dinklage} or homoclinic bifurcation~\cite{pre:jaman} depending on the plasma parameters.

Usually, dynamics of a periodically perturbed plasma depends on both the amplitude and frequency of the external perturbation that may be represented by ``Arnold tongue'' diagram. Though the dynamics of an excitable system depends on the frequency of the perturbation, it does not depend much on the perturbation amplitudes as long as the perturbation lifts the system at its excited state. This is because, the variation in perturbation amplitude does not affect much the system dynamics at excited state~\cite{physrep:linder}. Hence usual ``Arnold tongue'' diagram may not be obtained. Here, we have presented the dynamics of an excitable glow discharge plasma under subthreshold and suprathreshold periodic sawtooth forcing. The paper has been organized as follows: we present a brief description of the experimental setup in Sec.~\ref{section:setup} and autonomous system dynamics through which excitability was obtained in the glow discharge plasma has been discussed in Sec.~\ref{section:dynamics}. Experimental results has been presented and discussed in Sec.~\ref{section:result}. Finally, we conclude the results in Sec.~\ref{section:conclusion}.

\section{Experimental setup}
\label{section:setup}
\begin{figure} [hb]
\centering
\includegraphics[width=8.5 cm]{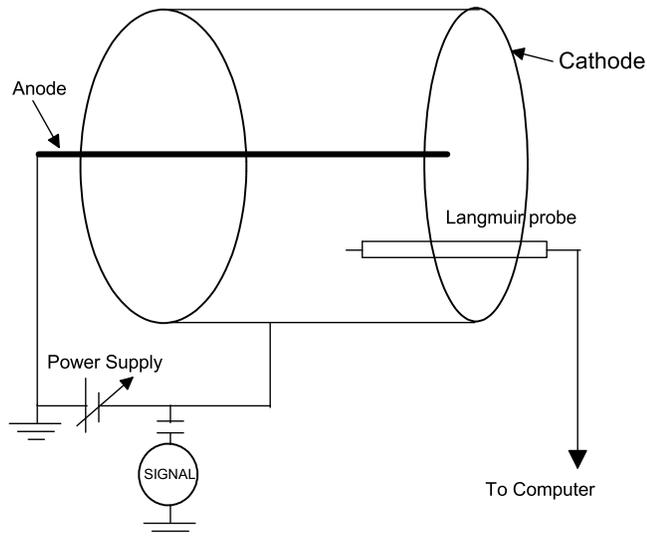}
\caption{Schematic diagram of the cylindrical electrode system of the glow discharge plasma. The probe is placed at a distance $l \approx12.5$ mm from the anode. }
\label{fig:setup}
\end{figure}

The experiments were carried out in a glow discharge argon plasma produced by a dc discharge in a cylindrical hollow cathode electrode system with a typical density and temperature $\approx10^7 cm^{-3}$ and $2-6$ eV respectively. The schematic diagram of the electrode system is shown in Fig.~\ref{fig:setup}. The electrode assembly was housed inside a vacuum chamber and the neutral pressure inside the chamber was controlled by a needle valve and the range of the gas pressure in these experiments was between 0.001 to a few mbar.  The discharge voltage ($0-999$ V) was applied between the cathode and the anode keeping the anode grounded. A signal generator (Fluke PM5138A) was coupled with the power supply for the experiments of periodically forced plasma. The main diagnostics was the Langmuir probe which was used to obtain the floating potential fluctuations. In these experiments the discharge voltage (DV) and pressure were the control parameters. Detail of the experiments can be found in Ref.~\cite{pre:jaman}.

\section{autonomous dynamics}
\label{section:dynamics}

\begin{figure*}[ht]
\centering
\includegraphics{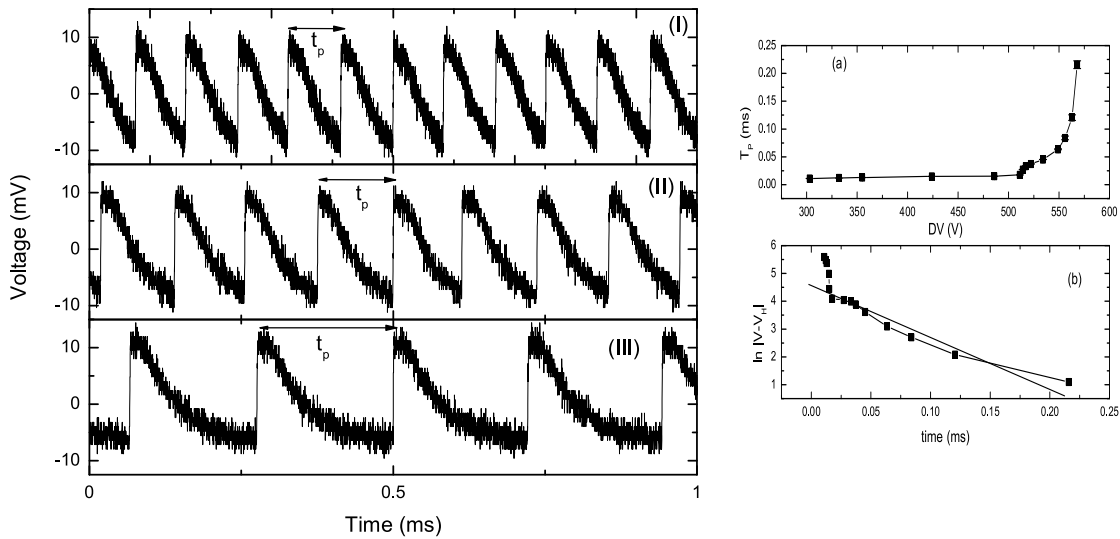}
\caption{Left panel: Time series showing that the period of relaxation oscillations increases with augmenting DV. Right panel :(a) Exponential increment of the time period (T) with DV and (b) $\ln|V-V_H|$ vs T curve can be fitted by a straight line indicating an underlying homoclinic bifurcation.}
\label{fig:homoclin}
\end{figure*}

By varying the neutral pressure, the breakdown of the gas occurs at different DV. At higher pressure (i.e., order of 1 mbar), with the initiation of discharge, a bright glow or spot was observed to form on the anode that was unstable initially, and was the source of the fluctuations. An interesting feature associated with the anode glow was the different types of oscillations in the floating potential at different pressures. At the initial stage, system showed irregular and relaxation type oscillations. Ions produced inside the anode glow due to collisions of the accelerated electrons across it with the neutrals, made the glow unstable~\cite{ppcf:sanduloviciu}, that was probably responsible for the appearance of these oscillations~\cite{pop:lee,pop:Klinger}.

Increasing the DVs, lead to an increase both in the amplitude and the time period of the oscillations and they become regular (period 1) at certain point of DV. Further increase in the DV modifies the oscillation profile and results in typical relaxation oscillations which change to excitable fixed point behavior at certain point. The DV at which these oscillations cease may be termed the bifurcation point ($V_H$) which depends also on the neutral pressure. The higher the neutral pressure, the higher the $V_H$. As the frequencies of the instabilities were within the ion plasma frequency (order of a few kHz) and the presence of anode glow and relaxation oscillations have been attributed to the presence of double layers, these instabilities could be related to ion acoustic instabilities and also anode glow related double layer~\cite{chaos:jaman,pop:jaman}.

The time period (T) of these relaxation oscillations increases  with increasing DV near the bifurcation point, and becomes infinite beyond the $V_H$, results in vanishing of the limit cycles. For larger values of DV, the autonomous dynamics exhibits a steady state fixed point behavior. Time traces from top to bottom in the left panel of Fig~\ref{fig:homoclin} depict this period lengthening of the  oscillatory behavior. A systematic analysis of the increment in the period (T), presented in Fig.~\ref{fig:homoclin}(a) [right panel], indicates that the autonomous dynamics undergoes a exponential slowing down. Consequently, the $\ln|V-V_H|$ vs T curve can be fitted by a straight line. The results of Fig~\ref{fig:homoclin} indicate that the system dynamics undergo a homoclinic bifurcation at $V_H$~\cite{pre:jaman}. The point $V_H$ divides the plasma into two distinct regions. The floating potential fluctuation exhibits relaxation oscillations on the one side of the $V_H$ and stable fixed point on the other side, which is a typical behavior of an excitable system. In the experiments, DV was so chosen that the system exhibited fixed point behavior and were perturbed by subthreshold and suprathreshold external sawtooth periodic signals.

\section{Results}
\label{section:result}
Dynamical behavior of the system under such perturbation has been presented in this section.

\subsection{Effect of subthreshold signal}
\label{subsection:subthreshold}

\begin{figure}[ht]
\centering
\includegraphics[width=8.5cm]{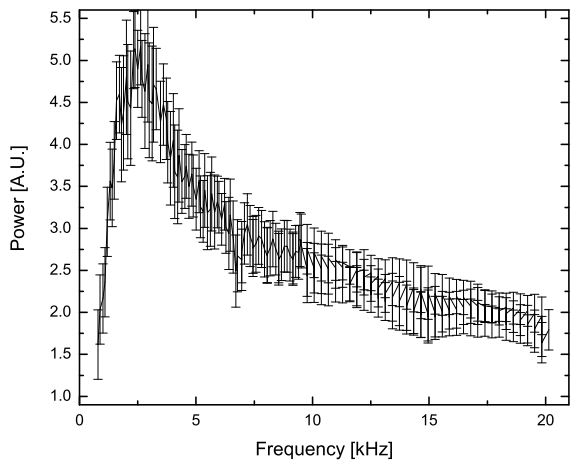}
\caption{Resonance curve of the subthreshold oscillations with frequency of the applied sawtooth signals. It peaks at system internal frequency around 3.2 kHz.}
\label{fig:subthreshold}
\end{figure}

For the study of the effect of subthreshold signal, the reference voltage $V_0$ was chosen such that $V_0>V_H$ and therefore the autonomous dynamics, by virtue of an underlying homoclinic bifurcation, exhibits steady state behavior. The discharge voltage V was thereafter perturbed, $V=V_0+S(t)$, where, the subthreshold sawtooth periodic signal $S(t)$ was so chosen that the voltage $V=V_0+S(t)<V_H$, i.e., the applied perturbation does not cause the system to cross over to the oscillatory regime. The subthreshold periodic sawtooth signal of variable frequency was generated using a Fluke PM5138A function generator. In this experiment, the frequency of the applied signal was varied from a few Hz to 20 kHz. When the frequency of the applied signal was around a few kHz, system showed subthreshold oscillations. Amplitude of these oscillations was increased with increase in the perturbation frequency that was recorded using an oscilloscope.

Figure~\ref{fig:subthreshold} shows the power of the subthreshold oscillations as a function of the frequency of the applied signals. It shows that initially the power of the subthreshold oscillations was minimum and increased with increase in frequency, and showed maximum value at frequency 3.2 kHz. Power again decreased with increase in perturbation frequency, hence system showed resonance at 3.2 kHz. As the frequency of the internal plasma oscillations which were basically the acoustic instabilities, was of the order of a few kHz in these parametric regions of plasma~\cite{pre:jaman,chaos:jaman,pop:jaman},  so system shows resonance at internal frequency. Therefore, these subthreshold oscillations may be useful to determine the system frequency. One important observation was that these oscillations were not observed far away from the bifurcation point. This indicates that the oscillations are the feature of an excitable dynamics~\cite{physrep:linder}. This approach may be useful to determine the frequency of internal oscillations of other excitable systems.

\subsection{Effect of suprathreshold signal}
\label{subsection:suprathreshold}

For the experiments on the effect of a suprathreshold forcing, the DV ($V_0$) was set such that the floating potential fluctuations exhibit fixed point behavior. In order to minimize the effect of parameter drift, a set point ($V_0$) quite far from the homoclinic bifurcation ($V_H$) was chosen. Subsequently, a suprathreshold sawtooth signal of fixed amplitude but variable frequency was superimposed on the DV. So in this case perturbation always crosses the threshold, i.e., $V=V_0+S(t)>V_H$. The frequency of the applied signals was varied from a few Hz to 20 kHz.
\begin{figure*}[ht]
\centering
\includegraphics[width=17cm]{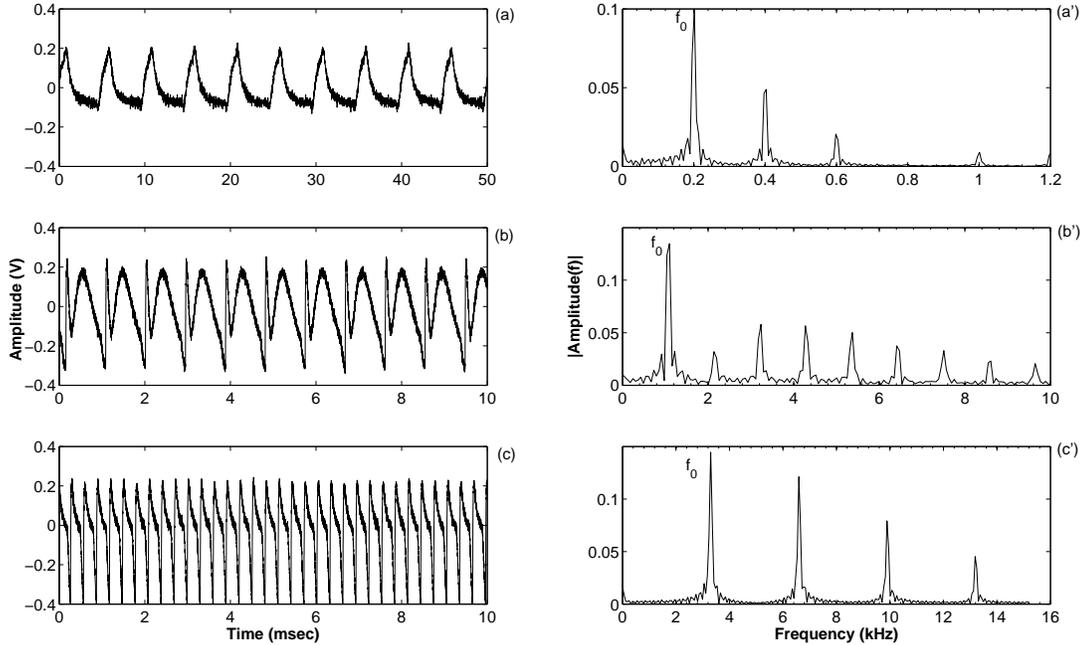}
\caption{Left panel: time series of plasma floating potential for periodic sawtooth perturbation: (a) at frequency 0.2
kHz; (b) at 1.072; and c) at 3.56 kHz, respectively, and their power spectra at (a$'$), (b$'$), and (c$'$) respectively. Figure(a) shows that system does not responds at frequency 0.2 kHz; figure(b) shows that the system its excitable behavior at 1.072 kHz and Fig.(c) shows 1:1 entrained state at 3.56 kHz.1:1 entrained state is also clear from the power spectrum [Fig.(c$'$)] also where internal frequency of the system coincides with external perturbation. }
\label{fig:entrainment1}
\end{figure*}

\begin{figure*}[ht]
\centering
\includegraphics[width=17cm]{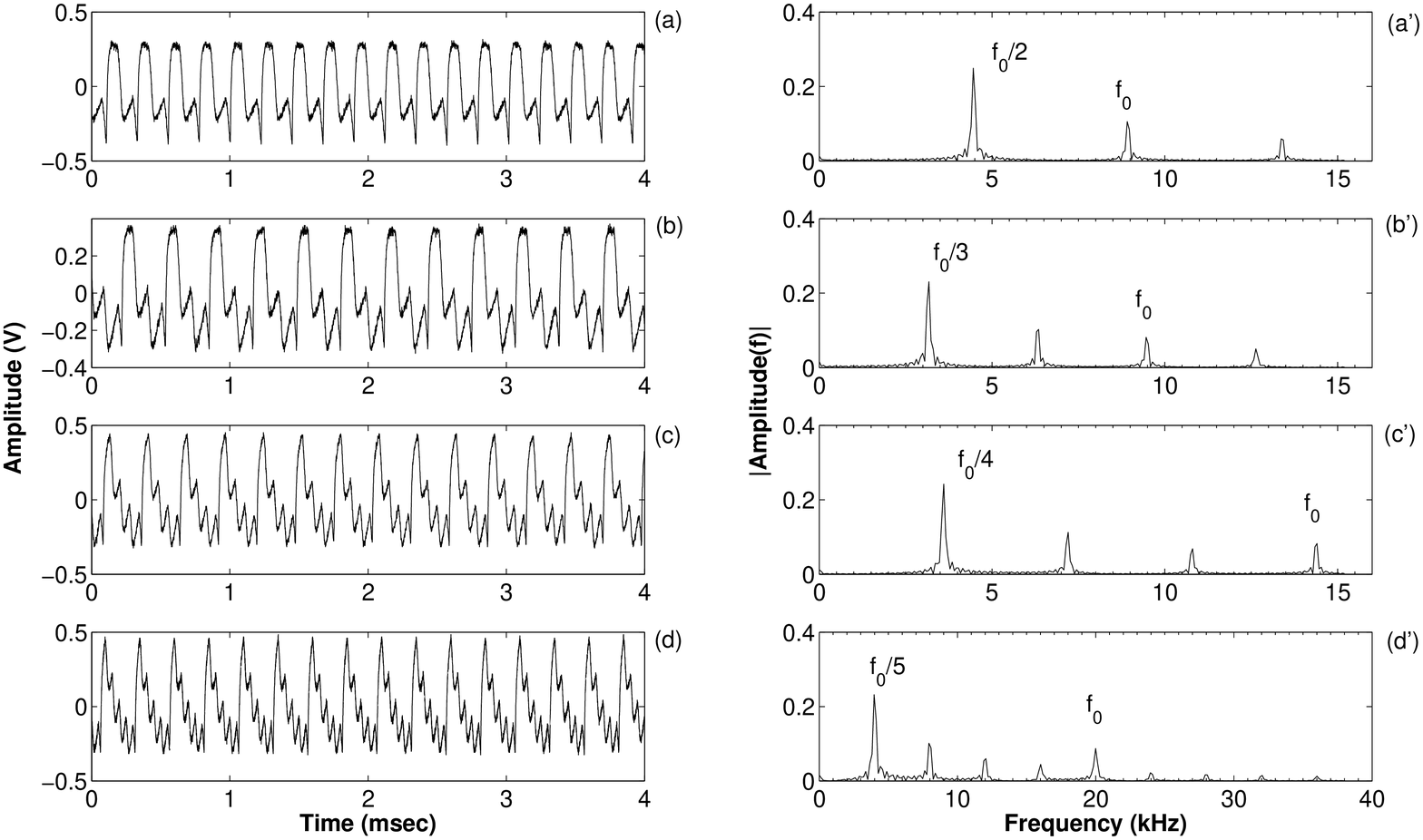}
\caption{Time series of plasma floating potential for periodic sawtooth perturbation: (a) at frequency 8.93 kHz; (b) at 9.48; c) at 14.38 kHz; and d) at 20.0 kHz respectively. Figure(a) shows 1:2 entrained; figure(b) shows 1:3 entrained
state; Fig.(c) shows 1:4 entrained state; and Fig.(d) shows 1:5 entrained state.}
\label{fig:entrainment2}
\end{figure*}

At low frequency, i.e., less than 1 kHz, though the amplitude of the applied signal is sufficient to cross the threshold, the system did not show any excitable dynamics. In these frequency, we obtained only the plasma distorted sawtooth signals as shown in Fig.~\ref{fig:entrainment1}(a) for perturbation frequency 0.2 kHz. We obtained perturbation invoked limit cycle oscillations at frequency 1.072 kHz as shown in Fig.~\ref{fig:entrainment1}(b). In this figure, the oscillations with larger and smaller time periods are the applied sawtooth signal and the limit cycle oscillations of the plasma system at excited state respectively. Oscillations with smaller time period were of the order of system's characteristic time scale (inverse of internal frequency). From Figs~\ref{fig:entrainment1}(a) and (b), it is also clear that the excitation mechanism of an excitable system depends not only on the perturbation amplitudes but also on the frequency.

Fig.~\ref{fig:entrainment1}(c) and (c$'$) show the system output and corresponding power spectrum when the perturbation frequency was 3.56 kHz. They clearly show that the system response is phased locked with the external periodic signal with rotation number 1:1. In case of phase locked states, rotation number (w) is defined as $W=f/f_0$, where $f$ and $f_0$ are the frequencies of external perturbation and system frequency, respectively~\cite{pre:klinger}. So when the oscillations frequency of the system was close to perturbation frequency ($f\approx f_0$) that is also clear from the frequency measurements presented in Subsec.~\ref{subsection:subthreshold}, 1:1 phase locking or entrainment phenomenon was obtained.  Figures~\ref{fig:entrainment1}(b) and (c) show that the periodic perturbations invoked limit cycle oscillations  at frequencies 1.072 and 3.56 kHz  respectively. The results shown Figures~\ref{fig:entrainment1}(b) cannot be taken as frequency entrainment phenomenon, as the phase is not locked with the external perturbation.  In this case, external perturbation lifts the system only at excited state and hence shows limit cycle oscillations. Whereas, at 3.56 kHz [Fig.~\ref{fig:entrainment1}(c)], external perturbations lifts the system at excited state as well as locks the phase of internal oscillations. From these observations it is clear that behavior of excitable plasma under external perturbation depends both on its excitable and internal dynamics.

\begin{figure}[ht]
\centering
\includegraphics[width=8.5cm]{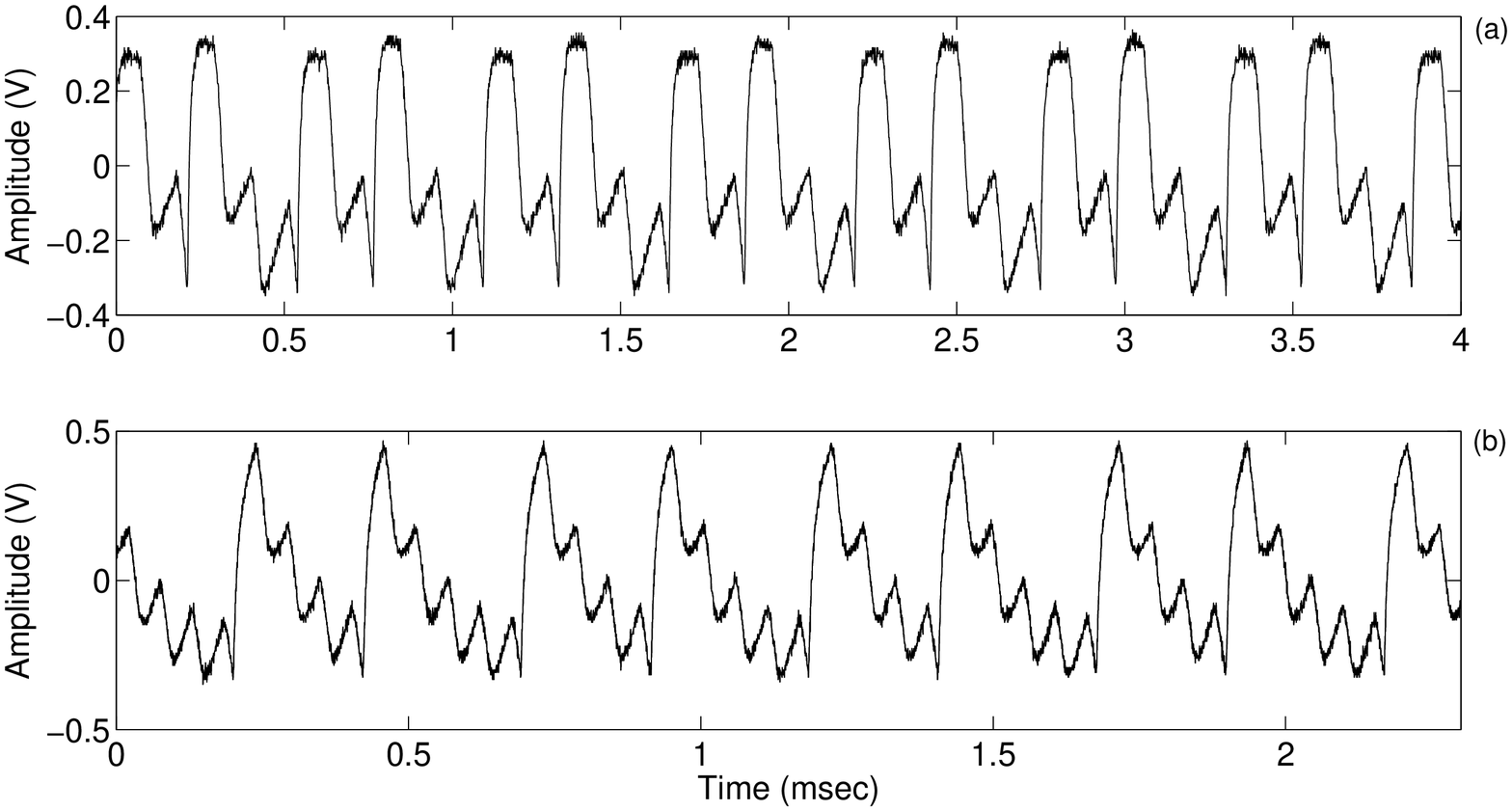}
\caption{Typical time series of plasma floating potential which show (a) intermittent entrainment between 1:2 and 1:3 at frequency 9.05 kHz; (b) intermittent entrainment between 1:4 and 1:5 at frequency 18.29 kHz.}
\label{fig:intermitten}
\end{figure}

When the perturbation frequency was increased to 8.93 kHz, system showed 1:2 harmonic frequency entrainment as shown in Fig.~\ref{fig:entrainment2}(a) and corresponding power spectrum shown in Fig.~\ref{fig:entrainment2}(a$'$). In Fig.~\ref{fig:entrainment2}(a$'$), $f_0$ represents the external perturbation and $f_0/2$ is the system oscillation at excited state. At perturbation frequencies 9.48, 14.38, and 20.0 kHz, we obtained 1:3, 1:4, and 1:5 entrainments or phase locked states, respectively. The entrainment phenomena is also clear from the times series shown Figs.~\ref{fig:entrainment2} (b), (c) and (d). These facts are also clear from their power spectra shown in Fig~\ref{fig:entrainment2} (b$'$), (c$'$) and (d$'$) respectively.

Between any two phase locked states, system responded such that it showed intermittent locked states.
Figures~\ref{fig:intermitten}(a) and (b) show the typical intermittent patterns between 1:2 and 1:3; and 1:4 and 1:5 respectively.

The above results show that the low frequency perturbations (less than its intrinsic frequency) only lift the system at excited state only, and hence system shows limit cycle oscillations which are not phase locked with perturbations. But for plasmas which exhibit van der Pol type behavior, show  2:1, 3:1, etc., subharmonic entrainments under low frequency~\cite{physcript:Klinger,pre:klinger}. In excitable plasma driving amplitude does not play not much role in the dynamics as long as the amplitude is sufficient to excite the system and hence  Arnold tongue diagram may not be obtained~\cite{physcript:Klinger,pre:klinger}.

\section{Conclusion}
\label{section:conclusion}
In this paper we have studied the dynamical behavior of excitable glow discharge plasma under the subthreshold and suprathreshold periodic forcing. System shows subthreshold oscillations under subthreshold forcing. These oscillations are the characteristics of the system dynamics and show resonance with external perturbations at the internal frequency of the system. Hence such resonance curve may be useful to characterize excitable plasma and this idea may also be applied in other excitable systems.

For the suprathreshold perturbation, system showed excitable or mode locking behavior depending on the perturbation frequency. Most simple frequency entrainment (1:1) phenomenon was obtained when the driver frequency was chosen close to system frequency. When the driver frequency was less that the system frequency, system showed natural excitable dynamics, whereas, for higher frequency, harmonic entrainments were observed. As the dynamics of an excitable system at excited state is almost independent of the amplitude of external perturbation, we did not get Arnold tongue behavior in the excitable region of glow discharge plasma.

\section*{Acknowledgement}
The authors acknowledge A. Bal and the Micro Electronic Division of SINP for their technical help during the experiments. One of the authors (MN) acknowledges support of Amit Apte at TIFR-CAM.


\begin{thebibliography}{100}


\bibitem{prl:abrahams} {R. H. Abrahams, Jr, E. J. Yadlowsky, and H. Lashinsky, Phys. Rev. Lett. \textbf{22}, 275 (1969).}
\bibitem{physcript:Klinger} {T. Klinger and A. Pie1, I. Axn\"{a}s and S. Torv\'en, Physica Scripta \textbf{56}, 70 (1997).}
\bibitem{pop:Rohde} {A. Rohde, A. Piel, and H. Klostermann, Phys. Plasmas \textbf{4}, 3933 (1997).}
\bibitem{ppcf:Gyergyek} {T Gyergyek, Plasma Phys. Control. Fusion \textbf{41}, 175 (1999).}
\bibitem{pop:Koepke} {M. E. Koepke, T. Klinger, F. Seddighi, and A. Piel, Phys. Plasmas \textbf{3}, 4421 (1996).}
\bibitem{pla:Klinger} {T. Klinger, A. Piel, F. Seddighi, and C. Wilke, Phys. Letts. A \textbf{182} 312 (1993).}
\bibitem{pop:Klinger} {T. Klinger, F. Greiner, A. Rohde, and A. Pie1, Phys. Plasmas \textbf{2}, 1822 (1995)}.
\bibitem{pre:klinger} {T. Klinger, F. Greiner, A. Rohde,  A. Piel, and M. E. Koepke, Phys. Rev. E, \textbf{52}, 4316 (1995).}
\bibitem{pop:dinklage} { A. Dinklage, C. Wilke and T. Klinger, Phys. Plasmas \textbf{6}, 2968 (1999)}.
\bibitem{pre:jaman} {Md. Nurujjaman, A.N.Sekar Iyengar, and P. Paramanda Phy. Rev. E \textbf{78}, 026406 (2008).}
\bibitem{prl:LinI} {Lin I and Jeng-Mei Liu, Phys. Rev. Lett. \textbf{74}, 3161 (1995)}.
\bibitem{prer:jaman} {Md. Nurujjaman, P. S. Bhattacharya, A. N. Sekar Iyengar, and Sandip Sarkar, Phy. Rev. E \textbf{80}, 015201 (R) (2009).}
\bibitem{prl:gang} {Hu Gang, T. Ditzinger, C. Z. Ning, and H. Haken, Phys. Rev. Lett. \textbf{71},
807 (1993)}.
\bibitem{prl:wiesenfeld} {Kurt Wiesenfeld, David Pierson, Eleni Pantazelou, Cris Dames,
and Frank Moss, Phys. Rev. Lett. \textbf{72}, 2125 (1994)}.
\bibitem{pre:strogatz} {W. J. Rappel and Steven Strogatz, Phys. Rev. E \textbf{50}, 3249 (1994)}.
\bibitem{prl:Giacomelli} {Giovanni Giacomelli, Massimo Giudici, Salvador Balle,
and Jorge R. Tredicce, Phys. Rev. Lett. \textbf{84}, 3298 (2000)}.

\bibitem{prl:pikovsky} {Arkady S. Pikovsky and J\"{u}rgen Kurths, Phys. Rev. Lett. \textbf{78},
775 (1997)}.
\bibitem{prl:locher} {M. L\"ocher,  D. Cigna,   and E. R. Hunt, Phys. Rev. Lett. \textbf{80}, 5212 (1998)}.
\bibitem{pre:bascones} {R. B\'ascones,  J. Garc\'ia-Ojalvo,   and J. M. Sancho,
Phys. Rev. E \textbf{65}, 061108 (2002)}.
\bibitem{rmp:sagues} {F. Sagu\'{e}s, J. M. Sancho, J. Garc\'ia-Ojalvo,
Rev. Mod. Phys. \textbf{79}, 829 (2007)}.
\bibitem{pre:dubbeldam} {Johan L. A. Dubbeldam, Bernd Krauskopf, and Daan Lenstra,
Phys. Rev. E \textbf{60}, 6580 (1999)}.
\bibitem{prl:stock} {N. G. Stocks, Phys. Rev. Lett. \textbf{84}, 2310 (2000).}
\bibitem{physrep:linder} { Lindner B, Garcia-Ojalvo J., Neiman A. and Schimansky-Geier L,
Phy. Rep.  \textbf{392}, 321 (2004).}
\bibitem{pre:santos1} { Gerando J. Escalera Santos, M. Rivera, M. Eiswirth, and P. Paramanda,
Phy. Rev.E \textbf{70}, 021103 (2004).}
\bibitem{pre:lindner} { Benjamin Lindner and Lutz Schimansky-Geier, Phy. Rev.E \textbf{61}, 6103 (2000).}
\bibitem{pre:lee} { Sang-Gui Lee, Alexander Neiman, and Seunghwan Kim, Phy. Rev.E \textbf{57}, 3292 (1998).}
\bibitem{pre:miyakawa} { Kenji Miyakawa and Hironobu Isikawa, Phy. Rev.E \textbf{66}, 046204 (2002).}
\bibitem{pre:Mendez} {Jorge Manuel Mendez, R. Laje, M. Giudici, J. Aliaga, and G. B. Mindlin, Phys. Rev. E
\textbf{63}, 066218 (2001)}.
\bibitem{pre:larotonda} {Miguel A. Larotonda, Alejandro Hnilo, Jorge M. Mendez, and Alejandro M. Yacomotti, Phys.
Rev.E \textbf{63}, 066218 (2001)}.

\bibitem{prl:kaplan} {Daniel T. Kaplan, John R. Clay, Timothy Manning, Leon Glass, Michael R. Guevara, and Alvin Shrier, Phys. Rev. Lett. \textbf{76}, 4074 (1996).}

\bibitem{J.theor.Biol.:clay} {John R. Clay and Alvin Shrier, J. theor. Biol. \textbf{197}, 207 (1999).}
\bibitem{r.soc:masoller} {Cristina Masoller, M. C. Torrent and Jordi Garc\'ia-Ojalvo, Phil. Trans. R. Soc. A \textbf{367}, 3255 (2009).}



\bibitem{chaos:Coullet} {P. Coullet, T. Frisch, J. M. Gilli, and S. Rica, Chaos \textbf{4}, 485
(1994)}.


\bibitem{ppcf:sanduloviciu}{M. Sanduloviciu and E. Lozneanu, plasa Phys. Control. Fusion \textbf{28}, 585 (1986)}.
\bibitem{pop:lee} { Hae June Lee and Jae Koo Lee, Phys. Plasmas \textbf{5}, 2878 (1998)}.
\bibitem{pop:jaman} {Md. Nurujjaman, Ramesh Narayanan, and A. N. Sekar Iyengar, Phys. Plasmas \textbf{16}, 102307 (2009).}
\bibitem{chaos:jaman} {Md. Nurujjaman, Ramesh Narayanan and A. N. Sekar Iyengar, Chaos \textbf{17}, 043121 (2007).}



\end{thebibliography}
\end{document}